\documentclass[aps,prl,twocolumn,groupedaddress,showpacs]{revtex4}

\usepackage{graphicx}
\begin{document}

\title{Photon-assisted tunneling in a Fe$_8$ Single-Molecule Magnet}

\author{L. Sorace$^{1,2}$, W. Wernsdorfer$^1$, C. Thirion$^1$, 
A.-L. Barra$^2$, M. Pacchioni$^3$, D. Mailly$^4$, B. Barbara$^1$}

\affiliation{
$^1$Laboratoire Louis N\'eel, associ\'e \`a l'UJF, CNRS, BP 166, 38042 Grenoble Cedex 9, France\\
$^2$Grenoble High Magnetic Field Laboratory, CNRS, BP 166, 38042 Grenoble Cedex 9, France\\
$^3$Dipartimento di Chimica and INSTM, Universit\`a di Firenze, 
Lastruccia 3, 50019 Sesto Fiorentino, Italy\\
$^4$LPN, CNRS, Route de Nozay, 91460 Marcoussis, France
}

\date{\today}

\begin{abstract}
The low temperature spin dynamics of a Fe$_8$ Single-Molecule Magnet was 
studied under circularly polarized electromagnetic radiation allowing us to establish 
clearly photon-assisted tunneling. This effect, while linear at low power, 
becomes highly non-linear above a relatively low power threshold. 
This non-linearity is attributed to the nature of the coupling of the sample to the thermostat.
These results are of great importance if such systems 
are to be used as quantum computers.

\end{abstract}


\maketitle

Molecular nanomagnets have attracted much interest in recent 
years both from experimental and theoretical 
point of 
view~\cite{Sessoli93b,Barra96,Thomas96,Friedman96,Aubin98,Prokofev98,WW_Science99,
Yoo_Jae00,Fernandez01,Hill02,Chudnovsky02,Fernandez02,Dressel03}. 
These systems are organometallic clusters characterized 
by a large spin ground state with a predominant 
uniaxial anisotropy. 
These features yield a pronounced barrier for magnetization 
reversal and therefore a hysteretic behavior below a blocking temperature. 
The quantum nature of these system makes 
them very appealing for phenomena occurring 
on the mesoscopic scale, i.e. at the boundary between classical 
and quantum physics.
 Up to now the most thoroughly investigated 
systems were a dodecanuclear mixed-valence 
manganese-oxo cluster with acetate ligands
(hereafter Mn$_{12}$-ac)~\cite{Sessoli93b} 
and an octanuclear iron(III) oxo-hydroxo cluster of 
formula [Fe$_8$O$_2$(OH)$_{12}$(tacn)$_6$]$^{8+}$ where tacn is a 
macrocyclic ligand, (hereafter Fe$_8$)~\cite{Barra96}, 
both having a ground spin state $S = 10$, but substantially 
differing in their anisotropy.
Indeed, while the first is essentially 
tetragonal with a barrier of 67 K, the latter is 
orthorhombic with a barrier of 25 K, leading to 
an enhancement of the ground-state tunneling rate compared to Mn$_{12}$-ac.

\begin{figure}
\begin{center}
\includegraphics[width=.45\textwidth]{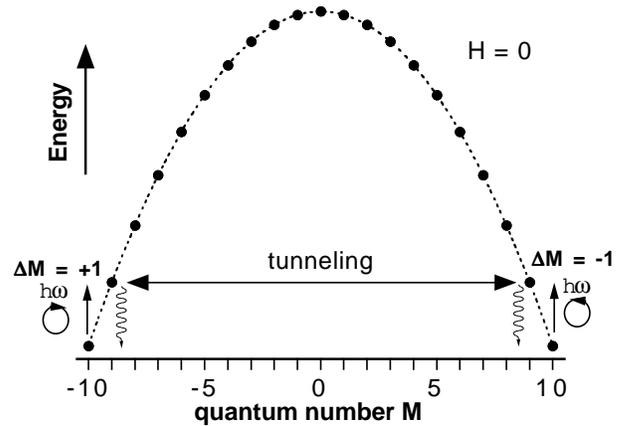}
\caption{Schematic representation of photon assisted tunneling. 
On irradiating a Fe$_8$ sample with a radiation of 
wavelength corresponding to the M = -10 to -9 splitting (vertical arrow), 
an enhancement of the fraction of molecules that tunnel 
from the first excited state is expected (horizontal arrow). 
The use of circularly polarized radiation allows 
to select only one side of the well and to distinguish 
between spin-phonon and spin-photon transitions.}
\label{schema}
\end{center}
\end{figure}

Recently it has also been proposed that molecular nanomagnets 
could be used as quantum computers by implementing 
GroverÕs algorithm~\cite{Leuenberger01}. For this to occur it 
is necessary to be able to generate an arbitrary superpositions 
of eigenstates of these systems. The suggested way to do 
this was through the use of multifrequency coherent magnetic 
radiation in the microwave and radiofrequency range. 
This would first introduce and amplify the desired phase 
for each $M$ state and this information could be 
finally read-out by standard magnetic resonance techniques. 
In this approach advantage is taken from the non-equidistance 
of the $M$ levels of the ground multiplet arising 
from the large axial anisotropy of these systems, 
which allows coherent populations of the different $M$ levels. 
A recent theoretical work pointed out that a very accurate 
control of pulse shape technique, both in amplitude, 
duration and choice of frequency is needed to fulfill the condition 
to design quantum computing devices in molecular nanomagnets~\cite{Zhou02}. 
In addition to such basic difficulties, we will see below that 
the microwave power cannot exceed a critical value above 
which non-linear effects occur. In particular the phonons 
associated with spin relaxation are crucially influencing 
the dynamics of the system. 

In order to investigate the feasibility of the proposed 
process any preliminary experiment should aim 
to understand the effects of microwave absorption 
on the spin dynamics of these systems at low temperature.
The main issues addressed in this letter concern:
(i) controlled increasing of excited state populations through the
absorption of microwave radiation, (ii) mechanism of 
photon-assisted tunneling, and (iii) subsequent heating 
occurring after relaxation. 

The measurements were performed on a new magnetometer, 
involving 10 $\times$ 10 $\mu$m$^2$ micro-Hall bars~\cite{Kent94,Kent00b}
in a dilution refrigerator equipped with three coils 
allowing to apply a field in any direction, 
at sweeping rates up to 1 T/s. 
Continuous radiation, in the FIR region, was generated 
by a couple of Gunn diodes (Radiometer Physics GmbH), 
working at fundamental frequencies of 95 GHz and 115 GHz, 
and equipped with double and triple harmonic generators 
and calibrated attenuators. We irradiated the sample 
using a 6 mm waveguide equipped with infrared filters 
in order to reduce heating. 
The circular polarization, induced by filters, was maximized around  
97\%. The study was performed on a 0.1 mm Fe$_8$ single 
crystal synthesized according to the standard procedure. 

\begin{figure}
\begin{center}
\includegraphics[width=.4\textwidth]{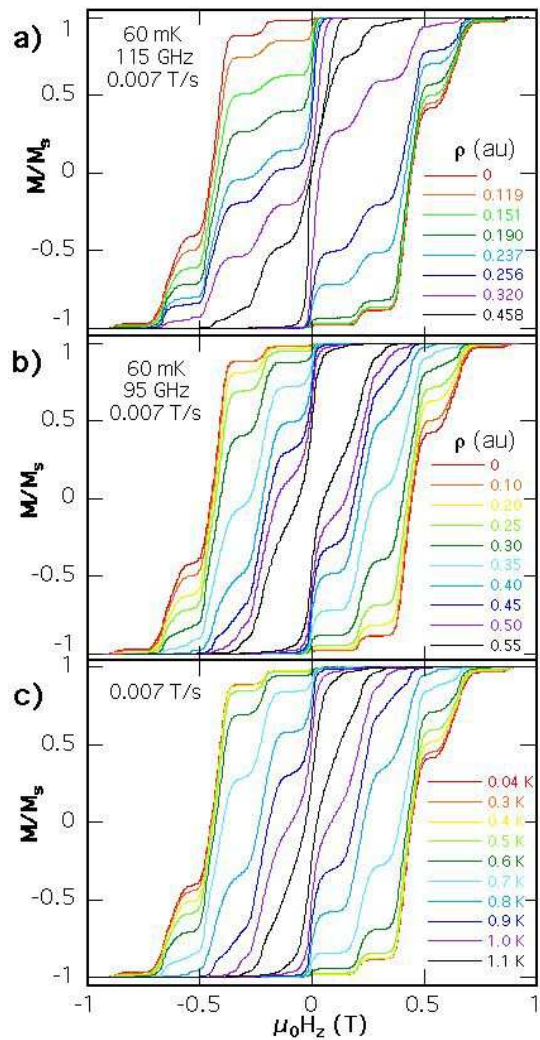}
\caption{{\it (color plot!)} Magnetic hysteresis loops of Fe$_8$ at a field sweep rate of 0.007 T/s
and at 60 mK under irradiation with microwaves at
(a) 115 GHz and (b) 95 GHz and for several microwave powers $\rho$.
The easy axis of the crystal is oriented along the applied field and 
perpendicular to the radiation oscillating magnetic field. 
The observed increasing of the tunneling rate at zero field, 
as a consequence of the absorption of photons induced by 
circularly polarized radiation, becomes evident by comparing 
the zero-field steps after positive or negative saturation. 
Comparing (a) with (b) clearly shows 
that the irradiation effect is much smaller for 95 GHz 
and resembles the thermal behavior presented in (c).}
\label{hyst}
\end{center}
\end{figure}
 
As schematically depicted in Fig. 1, microwave radiation
with a frequencies of 115 GHz corresponds to the energy
separation between the ground states $M = \pm S$ and 
the first excited states $M = \pm(S-1)$ of  Fe$_8$  
in zero applied magnetic field~\cite{Barra96,Hill02}. 
If the radiation is linearly polarized, 
the populations of the first excited states ($M = \pm(S-1)$) 
in both wells will be enhanced equally 
(equal transition probability for $\Delta M = \pm1$). On the opposite, 
the use of circular polarization has the advantage 
to distinguish between $\Delta M = +1$ (left polarization, 
$\sigma^-$ photons) or $\Delta M = -1$ (right polarization, 
$\sigma^+$ photons)~\cite{Abragam70}, and the population 
of only one of the two excited states will be 
enhanced (Fig. 1). 
An excess of tunneling from one well to the other is then expected. 
Therefore, circular polarization can help 
to distinguish between spin-phonons relaxation, 
and spin-phonons relaxation modified by the absorption of photons. 
The first equally affect the two sides of the barrier, 
i.e. the two branches of the hysteresis loop, 
while the second modifies the population of only one side of the barrier, 
i.e. one branch of the hysteresis loop. 
Any difference observed between the two branches 
of the hysteresis loop, has to be traced back to photon absorption. 

Fig. 2a shows the hysteresis loops of a Fe$_8$ single crystal 
with the easy axis parallel to the applied field, 
measured at 60 mK under irradiation. 
The tunneling transition near zero field is strongly 
enhanced for a radiation at 115 GHz. This is in agreement with a photon 
induced population transfer from $M = -10$ to $M = -9$, and agrees 
with earlier HF-EPR studies showing strong zero field 
absorption at about 116 GHz~\cite{Barra96,Hill02}. 
Fig. 2a also shows the expected asymmetry of the 
hysteresis loops in the presence of circularly polarized radiation. 
In particular, the height of the zero-field step 
(first tunnel resonance, $n = 0$), obtained when sweeping 
the field from negative saturation, is much less affected 
than when sweeping from positive saturation. Besides, 
the effect of a radiation at 95 GHz (Fig. 2b) with no level 
matching is quite comparable to that of temperature (Fig. 2c) 
(heating induced by large power irradiation). 
These results establish clearly 
that tunneling is assisted by photons for the matching frequency of 115 GHz. 
Similar qualitative effects were also observed in a field of 0.22 T, 
corresponding to the second tunnel resonance ($n = 1$). 

\begin{figure}
\begin{center}
\includegraphics[width=.5\textwidth]{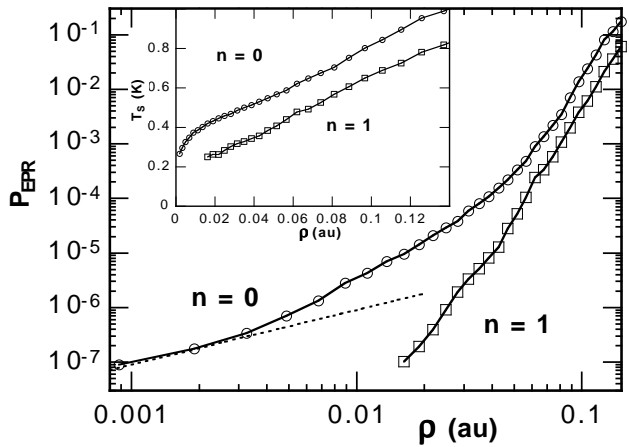}
\caption{Transition probability induced at 115 GHz as a function of the radiation power, 
measured by the Landau--Zener method at a sweeping  field rate of 0.14 T/s 
for the transition at $\mu_0H_z =$ 0 T (circles) and 0.22 T (squares). 
The dotted line shows the linear increase for small powers.
Inset: Spin temperature $T_S$ versus radiation power.}
\label{fig3}
\end{center}
\end{figure}

For the quantitative study of the power dependence 
of photons-induced tunneling,
we first cooled the 
sample from 5 K down to 0.04 K in a field 
of $H_z$~=~1.4~T yielding a positive 
saturated magnetization state.
Then, in the presence of continuous microwave radiation 
at 115 GHz and power $\rho$,
we swept the applied field at a constant rate 
over the zero field resonance transition and 
measured the fraction of molecules which 
reversed their spin. This procedure yields the 
total tunnel probability $P(\rho)$.
The photon induced tunneling probability 
$P_{EPR}(\rho)$ was evaluated from:
\begin{equation}
    P_{EPR}(\rho) = P(\rho) - n_{10}P_{\pm10}
\label{eq_P_EPR}
\end{equation}
where the ground state tunnel probability 
$n_{10}P_{\pm10}$ was measured in the absence of radiation 
~\cite{WW_EPL00}. The obtained $P_{EPR}(\rho)$ 
first increases linearly with $\rho$ and then becomes 
highly non-linear (Fig. 3).

\begin{figure}
\begin{center}
\includegraphics[width=.5\textwidth]{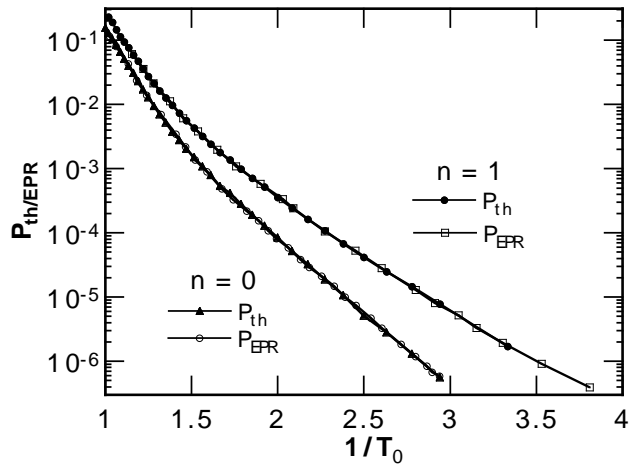}
\caption{Thermal transition probability $P_{th}(T) = P - 
n_{10}P_{\pm10}$ where the ground state tunnel probability 
$n_{10}P_{\pm10}$ was measured at 40 mK. 
The spin temperature $T_S$ in the inset of Fig. 3 
were determined by mapping $P_{EPR}(\rho)$ onto $P_{th}(T)$  
such as $P_{EPR}(\rho) = P_{th}(T_0 = T_S)$.}
\label{fig4}
\end{center}
\end{figure}

In order to give a simple explanation of the observed
behavior, we developed a scenario in which the spin temperature 
increases rapidly with the absorbed microwave power.  
We used the generalized master equation
formalism as described by Leuenberger and Loss~\cite{Leuenberger00}
but with photon induced transition probabilities $\Gamma_{\rm RF}$
between the energy states $M = \pm S$ and $\pm(S-1)$.
At low temperature ($T << 1$ K), it is sufficient to consider
only the six lowest levels (Fig. 1) 
with occupation numbers $n_M$ where
$M = \pm S, \pm(S-1), \pm(S-2)$. As initial condition
we have chosen positive saturation of the sample,
i.e. $n_{S} \approx N$, where N is the number of molecules.
We yield for small microwave power $\rho$ and $P_{EPR} << 1$:
\begin{equation}
P_{EPR} \approx \frac{2 \Gamma_{\rm RF}\Delta_{\pm9}}
                     {9 g \mu_{\rm B} \mu_0 dH_z/dt} 
   + \sum_{M = 9, 8} 
     \frac{\Delta_{\pm M} e^{-E_{10,M}/k_BT_S}}
          { \tau_M g \mu_{\rm B} M \mu_0 dH_z/dt}
\label{eq_PEPR}
\end{equation}
where  $\tau_M$ is the lifetime of the excited level $M$,
$E_{10,M}$ is the energy gap between the levels $10$ and $M$,
$\Delta_{\pm M}$ is the tunnel splitting of the levels $\pm M$,
and $T_S$ is the spin temperature. All constants of
this model were found by measuring 
the overall transition rate via excited spin levels $P_{th}(T)$ 
due to thermal activation, as described in~\cite{WW_EPL00}.
The spin temperature $T_S$ were determined by mapping
$P_{EPR}(\rho)$ onto $P_{th}(T)$ 
such as $P_{EPR}(\rho) = P_{th}(T_0 = T_S)$ (Fig. 4).
We obtained a nearly linear variation of $T_{S}$ versus $\rho$
(inset of Fig. 3).

The energy transfer from the spin system to the thermostat 
involves contributions of the electromagnetic 
radiation bath, the phonon bath and the spin bath. 
However, at low temperatures ($T < 1$ K) the phonon specific 
heat is vanishingly small and the photon bath 
completely negligible. The specific heat $C_S$ of the spin bath is 
greatly influenced by weak dipolar fields 
giving a flat distribution of Schottky  anomalies 
with a nearly constant specific heat~\cite{Gomes98,Mettes01}. 
The electromagnetic power absorbed by the spin system, 
$dW/dt = \hbar\omega(1 - n_{S-1}/N)\Gamma_{\rm RF} \approx \hbar\omega\Gamma_{\rm RF}$~\cite{Abragam70},  
passes to the thermostat via spin--lattice and spin--spin 
relaxation. The effect of the latter should 
be more important due to their larger specific heat. 
Neglecting the phonon specific heat, the heat transfer equation can be written:
\begin{equation}
dW/dt  =  C_SdT/dt   +  C_S(T_S - T_0)/\tau_S  
\label{heat_eq}
\end{equation}
where $T_0$ is the cryostat temperature and
$\tau_S$ is a diffusion time for magnetic excitations. 
At equilibrium $(dT/dt = 0)$,  the spin-bath temperature 
is therefore $T_S = T_0 + \hbar\omega\Gamma_{\rm RF}\tau_S/C_S$.
Because $C_S$ is almost temperature independent between
0.3 and 0.8 K~\cite{Gomes98,Mettes01}, $T_S$ increases 
nearly linearly with the microwave power which
is in good agreement with our observation (inset of Fig. 3).

Finally, the microwave absorption at 95 GHz and for $n = 1$
at 115 GHz might be due to spin--spin interactions 
which broaden strongly the energy levels. Such a
broadening has already been observed by EPR linewidth
measurements~\cite{Hill02} and spin--spin cross--relaxation
measurements~\cite{Giraud01,WW_PRL02}.

In conclusion, the use of circularly polarized microwaves 
allowed us to show for the first time the phenomenon of 
photon-assisted tunneling in magnetism, using a single molecule magnet 
Fe$_8$. In accordance with the selection rules 
for EPR spectroscopy~\cite{Abragam70}, circularly polarized 
radiation promotes the transition $M$ = -10 to -9 with $\Delta M$ = +1, 
giving an effect of magnetic dichroism at millimeter wavelengths. 
At lowest microwave powers, 
the tunnel probability increases linearly with the power, 
whereas at higher powers we enter in a non-linear regime
resulting from an increase of the spin temperature $T_S$. 

This work was supported by the 
European Union TMR network MOLNANOMAG, 
HPRN-CT-1999-0012.


\begin{thebibliography}{10}

\bibitem{Sessoli93b}
R. Sessoli, H.-L. Tsai, A.~R. Schake, S. Wang, J.~B. Vincent, K. Folting, D.
  Gatteschi, G. Christou, and D.~N. Hendrickson, J. Am. Chem. Soc. {\bf 115},
  1804  (1993).

\bibitem{Barra96}
A.-L. Barra, P. Debrunner, D. Gatteschi, Ch.~E. Schulz, and R. Sessoli,
  EuroPhys. Lett. {\bf 35},  133  (1996).

\bibitem{Thomas96}
L. Thomas, F. Lionti, R. Ballou, D. Gatteschi, R. Sessoli, and B. Barbara,
  Nature (London) {\bf 383},  145  (1996).

\bibitem{Friedman96}
J.~R. Friedman, M.~P. Sarachik, J. Tejada, and R. Ziolo, Phys. Rev. Lett. {\bf
  76},  3830  (1996).

\bibitem{Aubin98}
S.~M.~J. Aubin, N.~R. Dilley, M.~B. Wemple, G. Christou, and D.~N. Hendrickson,
  J. Am. Chem. Soc. {\bf 120},  839  (1998).

\bibitem{Prokofev98}
N.V. Prokof'ev and P.C.E. Stamp, Phys. Rev. Lett. {\bf 80},  5794  (1998).

\bibitem{WW_Science99}
W.Wernsdorfer and R. Sessoli, Science {\bf 284},  133  (1999).

\bibitem{Yoo_Jae00}
J. Yoo, E.~K. Brechin, A. Yamaguchi, M. Nakano, J.~C. Huffman, A.L. Maniero,
  L.-C. Brunel, K. Awaga, H. Ishimoto, G. Christou, and D.~N. Hendrickson,
  Inorg. Chem. {\bf 39},  3615  (2000).

\bibitem{Fernandez01}
J.~J. Alonso and J.~F. Fernandez, Phys. Rev. Lett. {\bf 87},  097205  (2001).

\bibitem{Hill02}
S. Hill, S. Maccagnano, Kyungwha Park, R.~M. Achey, J.~M. North, and N.~S.
  Dalal, Phys. Rev. B {\bf 65},  224410  (2002).

\bibitem{Chudnovsky02}
E.~M. Chudnovsky and D.~A. Garanin, Phys. Rev. Lett. {\bf 89},  157201  (2002).

\bibitem{Fernandez02}
Julio~F. Fernandez, Phys. Rev. B {\bf 66},  064423  (2002).

\bibitem{Dressel03}
M. Dressel, B. Gorshunov, K. Rajagopal, S. Vongtragool, and A.~A. Mukhin, Phys.
  Rev. B {\bf 67},  060405  (2003).

\bibitem{Leuenberger01}
M.~N. Leuenberger and D. Loss, Nature {\bf 410},  789  (2001).

\bibitem{Zhou02}
B. Zhou, R. Tao, S.Q. Shen, and J.~Q. Liang, Phys. Rev. A {\bf 66},  010301
  (2002).

\bibitem{Kent94}
A.D. Kent, S. von Molnar, S. Gider, and D.D. Awschalom, J. Appl. Phys. {\bf
  76},  6656  (1994).

\bibitem{Kent00b}
L. Bokacheva, A.D. Kent, and M.A. Walters, Phys. Rev. Lett. {\bf 85},  4803
  (2000).

\bibitem{Abragam70}
A. Abragam and B. Bleaney, {\em Electron paramagnetic resonance of transition
  ions} (Clarendon Press, Oxford, 1970).

\bibitem{WW_EPL00}
W. Wernsdorfer, A. Caneschi, R. Sessoli, D. Gatteschi, A. Cornia, V. Villar,
  and C. Paulsen, EuroPhys. Lett. {\bf 50},  552  (2000).

\bibitem{Leuenberger00}
M.~N. Leuenberger and D. Loss, Phys. Rev. B {\bf 61},  12200  (2000).

\bibitem{Gomes98}
A.M. Gomes, M.A. Novak, R. Sessoli, A. Caneschi, and D. Gatteschi, Phys. Rev. B
  {\bf 57},  5021  (1998).

\bibitem{Mettes01}
F.~L. Mettes, F. Luis, and L.~J. de~Jongh, Phys. Rev. B {\bf 64},  174411
  (2001).

\bibitem{Giraud01}
R. Giraud, W. Wernsdorfer, A.~M. Tkachuk, D. Mailly, and B. Barbara, Phys. Rev.
  Lett. {\bf 87},  057203  (2001).

\bibitem{WW_PRL02}
W. Wernsdorfer, S. Bhaduri, R. Tiron, D.~N. Hendrickson, and G. Christou, Phys.
  Rev. Lett. {\bf 89},  197201  (2002).

\end{thebibliography}
\end{document}